# Iterative multi-scale method for estimation of hysteresis losses and current density in large-scale HTS systems

Edgar Berrospe-Juarez[1], Víctor M R Zermeño[2], Frederic Trillaud[3] and Francesco Grilli[2]

[1]Postgraduate School of Engineering, National Autonomous University of Mexico, Mexico

[2]Karlsruhe Institute of Technology, Germany

[3]Institute of Engineering, National Autonomous University of Mexico, Mexico

email: eberrospej@iingen.unam.mx

**Abstract:** In recent years, commercial HTS materials have gained an increasing interest for their use in applications involving large-scale superconductor systems. These systems are typically made from hundreds to thousands of turns of conductors. These applications can range from power engineering devices like power transformers, motors and generators, to commercial and scientific magnets. The available analytical models are restricted to the analysis of individual tapes or relatively simple assemblies, then it is not possible to apply these models to the study of large-scale systems and other simulation tools are required. Due to the large number of turns, the simulations of a whole system can become prohibitive in terms of computing time and load. Therefore, an efficient strategy which does not compromise the accuracy of calculations is needed. Recently, a method, based on a multi-scale approach, showed that the computational load can be lowered by simulating, in detail, only several significant tapes from the system. The main limitation of this approach is the inaccuracy of the estimation of the background magnetic field, this means the field affecting the significant tapes produced by the rest of the tapes and by external sources. To address this issue, we consider the following two complementary strategies. The first strategy consists in the iterative implementation of the multi-scale method. The multi-scale method solves itself a dynamic problem, the iterative implementation proposed here is the iterative application of the multi-scale method, and a dynamic solution is obtained at each iteration. The second strategy is a new interpolation method for current distributions. With respect to conventional interpolation methods, a more realistic current density distribution is then obtained, which allows for a better estimation of the background magnetic field, and consequently, a better estimation of the hysteresis losses. In contrast with previous works, here we do not focus only on the estimation of the hysteresis losses, but also the estimation of the current density distribution is addressed. This new method is flexible enough to simulate different sections of the system with a better level of detail while providing a faster computational speed than other approaches. In order to validate the proposed method, a case study is analyzed via a reference model, which employs the $H$-formulation of Maxwell's equations and includes all the system's tapes. The comparison, between the reference model and the iterative multi-scale model, shows that the computation time and memory demand are greatly reduced. In addition, a very good agreement with respect to the reference model, both at a local and global scale, is achieved.

Keywords: large-scale superconductor systems, hysteresis losses, HTS superconducting magnets.

# 1. Introduction

The development of High Temperature Superconductor (HTS) materials has led to the commercialization of conductors with high current capacities that allow its application in superconductor power devices such as fault current limiters, generators, and SMES [1]; and also, in tools for medical al chemistry applications (MRI, NMR) [2]. Even though the prices of current HTS conductors make these devices still expensive, there is a sustained progress in the characteristics of the materials [3]. The high number of conductor turns in those devices permits to classify them as large-scale superconducting systems, this classification also includes other systems, such as high field magnets. To ensure a safe operation of these devices, their design should consider transient effects that may arise from changes in the external magnetic field and in the transport current, these changes can occur together, but can also occur separately. Indeed, during these changes, hysteresis losses are generated, which leads to temperature rises and potentially to the loss of the superconducting state. In any case, it is always important to assess the risks associated with the technology [4].

It is necessary to understand the electromagnetic properties and phenomena. The superconducting materials can be described by means of a power-law relating the electrical field $E$ to the current density $J$ [5]. To improve the accuracy of the model, it is often necessary to include the dependence of the critical current density ($J_c$) and the $n$-value on the magnetic flux density ($B$) [6]. These nonlinearities combined with the size of the large-scale systems turns the calculation of hysteresis losses into a cumbersome task, for which the simulation time and memory requirement can quickly become prohibitive [7]. This is a tremendous setback when the optimization of the device design is addressed by means of parametric simulations. Although there are some analytical models that provide a simpler closed form solution, these models are limited to very simple case studies. Other limitations include the fact of being limited to the critical state, and the difficulty of inserting complex $J_c$ dependence on $B$. For example, the models presented in [8-10] apply just for one single conductor, whereas the models presented in [11-14] apply for conductor stacks under restrictive conditions. A detailed compilation of analytical results is presented in [15].

Numerical methods such as the Finite Element Method (FEM) are very useful tools because they allow building detailed models that are able to estimate locally the electromagnetic quantities in the superconductor. In the context of large-scale systems, where the main problem is the large number of turns, a detailed comparison of three models for the estimation of hysteresis losses can be found in [16]. In that publication, a particular attention was paid to the homogenization method [17] and to the further development of the multi-scale method [18, 19]. Recently, an efficient FEM model based on the T-A formulation of Maxwell's equations has been discussed in [20, 21] for its use in large-scale systems as well. An alternative strategy is the minimum magnetic energy variation (MMEV) method, described in [22, 23].

Our interest revolves around the multi-scale method, since it allows reducing the size of the problem by analyzing in detail only several significant tapes. The accuracy of the method depends on how accurate the background magnetic field can be evaluated throughout the system [16]. As the background magnetic field is computed from the current density distribution, the latter should be accurately assessed over space and time. In this manuscript, we understand current density distribution ($J$ distribution) as the function associating a current density value to every point inside the tapes at every time step. The iterative method proposed in this manuscript initiates with a uniform *J* distribution in every tape and applies the multi-scale method to find a dynamic solution, e.g. a solution at every time step. The method is applied iteratively to find a new and more accurate dynamic solution by recomputing a better *J* distribution in every analyzed tape until a convergence criterion is fulfilled. As a complement to this strategy, we also propose a new interpolation method to approximate the *J* distributions in the tapes that are not analyzed in detail. The interpolation method is based on the inverse cumulative density function (ICDF) interpolation method, presented in [24]. In contrast with previous works, here the attention is focused not only on the estimation of the hysteresis losses, but also on the accurate estimation of the current density distribution. The accurate estimation of the background field and the hysteresis losses occur as a consequence of the accurate estimation of the *J* distribution.

The case study considered in this manuscript is a coil wound with HTS tapes, but the multi-scale method is not limited to this kind of systems, as it was demonstrated in [19] where this method is applied to analyze a coil wound with multi-filament $MgB_2$ conductors. Also, it is possible to apply the iterative multi-scale method to Low Temperature Superconductor (LTS) systems, the LTS materials can also be described by means of a power-law relating *E* and *J*, but the particulars of each system should be considered. The ICDF interpolation as well as most of the results in this manuscript are just valid for systems made with tapes with a thin superconducting layer. It is a well-known fact that the $J_c$ and the *n*-value may vary significantly along the conductor length [25, 26]. These variations can drastically affect the results. The case study considered here does not take into account these variations. The iterative multi-scale method provides the possibility to assign different properties to different sections of the system, however including a more realistic variation of the properties is not possible. Nevertheless, it is possible to simulate different cases for the best and worst possible scenario.

This manuscript is organized as follows. Section 2 contains the description of the multi-scale method and the detailed explanation of how this method was improved and transformed into an iterative method. The description of the case study and its reference model is presented in section 3. This reference model allows assessing the accuracy of the proposed improvements. The multi-scale model of the tested system and the simulation results are presented in sections 4. In this manuscript the simulations were conducted considering a sinusoidal transport current, but the method is not limited to this kind of transport currents. Also, it is possible to conduct simulations under transient external magnetic fields. Section 5 presents the conclusions of the work. Finally, the appendix A describes the new ICDF interpolation method for *J* distributions.

## 2. Iterative multi-scale method

**Multi-scale method**
The multi-scale method, as presented in [16, 18] relies on two 2D submodels and solves Maxwell's equations using FEM. The first submodel is an *A*-formulation magnetostatic model of the full coil that includes all the tapes with their actual geometry and is called *coil submodel*. The constitutive equation of the coil submodel is

$$\nabla \times \nabla \times \boldsymbol{A} = \mu_0 \boldsymbol{J}_a, \qquad (1)$$

where $\boldsymbol{J}_a$ is the applied current density. The other submodel is an *H*-formulation model of a single tape and is called *single tape submodel*, and its constitutive equation is

$$\nabla \times \rho \nabla \times \mathbf{H} = -\mu_0 \frac{\partial (\mathbf{H})}{\partial t}. \qquad (2)$$

The process to compute the losses has two steps and the data flow is unidirectional from the coil submodel to the single tape submodel. The first step is to use the coil submodel to estimate the background magnetic field strength *H* in every tape. Subsequently, the magnetic field strength along the boundary of some significant tapes, called *analyzed tapes*, is exported to the single tape submodel as a time-varying Dirichlet boundary condition. The hysteresis losses are calculated in these significant tapes. Then, the losses in the *non-analyzed tapes* are obtained by interpolation without the need of computing their *J* distribution. The breaking up of the problem into several smaller problems reduces the computational burden and more importantly, allows the parallelization of the problem.

The main limitation of the multi-scale method described above is the estimation of the background magnetic field, which, in turn, depends on the *J* distribution in all the system's tapes. In [16], the background field is approximated using two different approaches. The first approach is to consider a uniform $J_0$ distribution, this means a *J* distribution which assigns the same current density value for every point inside the tapes. It will be shown later that the assumption may result in significant errors. The second approach is to consider the *J* distribution produced by an infinite array, which provides a better approximation of the actual *J* distribution. As expected, the losses estimated using the second approach are more accurate due to a better estimation of the magnetic field distribution across the system.

**Iterative multi-scale method**

The original multi-scale method allows the data flow from the coil submodel into the single tape submodel. As discussed previously, the method accuracy mainly depends on the accurate estimation of the local magnetic field. Therefore, the proposed solution to improve the estimated background magnetic field is to implement an iterative procedure, as shown in figure 1. The new approach allows the data to flow from the single tape submodel to the coil submodel, and then back to the single tape submodel. This feedback loop allows computing a new and more accurate *J* distribution of the analyzed tapes with every iteration.

To initialize the iterative procedure, the coil submodel is used to estimate the magnetic field in the whole coil from a set of uniform $J_0$ distributions. The coil submodel is a magnetostatic model, thus this model is run as many times as time steps are needed. To each time step corresponds a different uniform $J_0$ distribution which, in turns, correspond to a different transport current. The magnetic fields along the boundary of the analyzed tapes for every time step are collected to build time dependent boundary conditions, which are then exported to the single tape submodel. Here, the *J* distributions in the analyzed tapes at every time step are computed and, by interpolation, the *J* distributions in the non-analyzed tapes are estimated. After exporting this new dynamic *J* distribution for all the tapes to the coil submodel, a new estimate for the magnetic field is calculated. At this point the first estimates for both the *J* distribution and the magnetic field have been calculated. The process is then repeated to obtain better estimates for both quantities. To exit from the iterative loop, the $J$ distribution of the present iteration is compared with the one at the previous iteration. Thus, if the error ( $Er$ ) between the $J$ of the present iteration and the one of the previous iteration is smaller than a user-defined criterion ($\varepsilon$), then the process is finished. The error, at the iteration $k$, is computed as

$$Er_k = \frac{\sqrt{\sum_{i=1}^{l}(J_i^{k-1}-J_i^k)^2}}{\sqrt{\sum_{i=1}^{l}(J_i^{k-1})^2}} \quad . \tag{3}$$

For a determined iteration $k$, $J^k$ is a vector containing the evenly sampled $J$ distributions for all the analyzed tapes, at every time step concatenated one behind the other. While $J^{k-1}$ contains the samples for the previous iteration, both vectors contain $l$ samples. As long as the single tape submodel considers just one element along the HTS layer thickness, the $J$ distributions at every time step are function of a single spatial variable, along the width of the tape. The error $Er$ allows comparing the $J$ distributions at all time steps. This is important because a dynamic solution is found at every iteration and the dynamic $J$ distributions of the iteration $k$ is compared with the dynamic $J$ distributions of the iteration $k$-1.

The hysteresis losses in the analyzed tapes are computed within the single tape submodel, while the losses of the non-analyzed tapes are interpolated from the losses in the analyzed tapes. The interpolation method used to approximate the losses in the non-analyzed tapes is a Piecewise Cubic Hermite Interpolating Polynomial method [27]. Figure 1 shows the flowcharts of the multi-scale and the iterative multi-scale method, to facilitate the comparison between both methods.

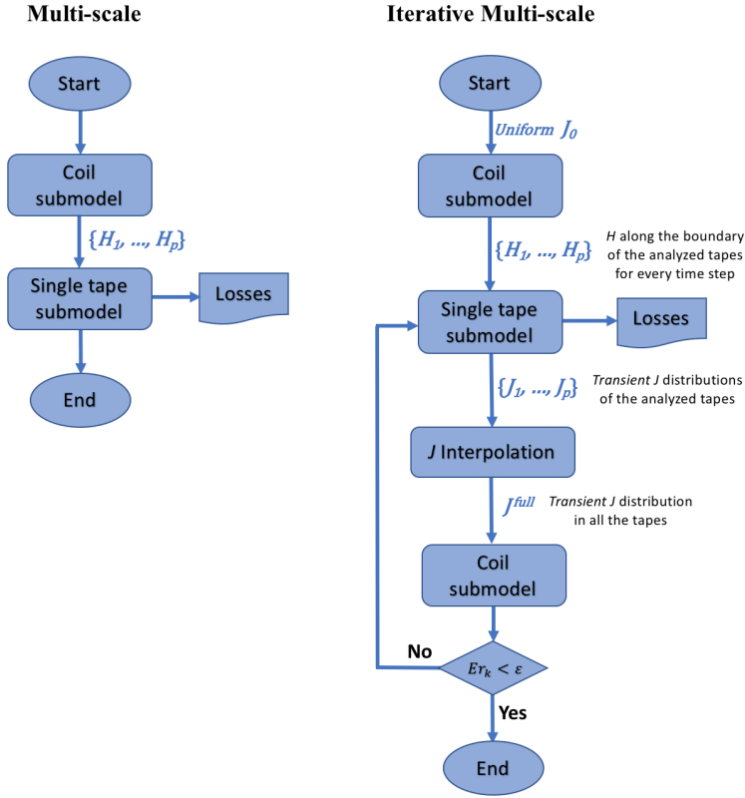

**Figure 1.** Flowcharts for the multi-scale and iterative multi-scale methods. Both processes begin with a uniform $J_0$ distribution, the one which assigns the same $J$ value for every point inside the tapes. The $H$ field at the boundary of the $p$ analyzed tapes is exported form the coil submodel to the single tape submodel. Then, the $J$ distributions from the analyzed tapes, referred as $\{J_1, …, J_p\}$ are used to approximate by interpolation the distribution in the non-analyzed tapes. The $J$ distribution containing the distributions of the analyzed and non-analyzed tapes is referred as $J^{full}$. The $J^{full}$ is exported to the coil submodel to compute a new background field.

## 3. Case study and reference model

**Case study**

The case study used in this manuscript to compare the models is a stack of series-connected HTS pancake coils forming a solenoidal magnet, the pancakes are wound with (RE)BCO coated conductors. It is assumed that the pancake coils have the same number of turns and each turn has the same dimensions. The magnet has 10 pancakes, each with 80 turns. The relevant parameters of the coil are summarized in table 1.

**Table 1.** Case study parameters.

| Parameter | Value |
| --- | --- |
| Inner radius | 20 mm |
| Outer radius | 40 mm |
| Height | 44.5 mm |
| Pancakes | 10 |
| Turns per pancake | 80 |
| HTS layer width | 4 mm |
| HTS layer thickness | 1 µm |
| Unit cell width | 4.45 mm |
| Unit cell thickness | 250 µm |

**Reference model**

The reference model is a 2D axisymmetric FEM model that includes all the tapes. It uses the *H*-formulation [28] of the Maxwell's equations. The constitutive equations have already been presented in [29] and [30]. Due to the symmetries, it is possible to employ adequate boundary conditions and model just one quarter of the cross-section of the coil. The electrical resistivity of the HTS material is modeled by the so-called *E-J* power-law [5],

$$\rho_{HTS} = \frac{E_c}{J_c(\boldsymbol{B})} \left| \frac{\boldsymbol{J}}{J_c(\boldsymbol{B})} \right|^{n-1}. \tag{2}$$

The anisotropic dependence of the $J_c$ on the magnetic field is given by a modified Kim's relation [31],

$$J_C(\boldsymbol{B}) = \frac{J_{c0}}{\left(1 + \frac{\sqrt{k^2 B_\parallel^2 + B_\perp^2}}{B_0}\right)^\alpha}, \tag{3}$$

where $B_\perp$ and $B_\parallel$ are the magnetic field components perpendicular and parallel to the flat surface of the tape, respectively. Although, it is known that the *n*-value in (2) depends on the magnetic field and it is possible to implement this dependence (an example with a similar numerical model can be found in [32]), here it is considered constant in order to make the comparison simpler. The field level of this case study justifies the assumption of a constant *n*-value, moreover, the variations on the *n*-value play a negligible effect on the hysteresis losses [6]. The parameters in (2) and (3), are reported in table 2, this set of parameters represents a (RE)BCO coated conductor [33]. The HTS layers are considered to be surrounded with a medium, having a resistivity equal to 1 Ωm as proposed in [17]. Also, the permeability of the full system is chosen to be equal to the permeability of the vacuum $\mu_0$, since there are no magnetic materials in the system.

Table 2. HTS parameters.

| Quantity | Value |
|---|---|
| $E_c$ | 1e-4 Vm$^{-1}$ |
| $n$ | 25 |
| $J_{c0}$ | 4.5e10 Am$^{-2}$ |
| $B_0$ | 0.03 T |
| $k$ | 0.2 |
| $\alpha$ | 0.6 |

The regions encompassing one single HTS layer and its surrounding medium are called unitary cells, as shown in figure 2. The mesh in every unitary cell is structured. The HTS layer has 1 element across its thickness and 100 elements along its width. The mesh is graded with increasing number of elements at the extremities of the HTS layer, because the penetration of the magnetic field into the tape starts from there. The mesh used for the unitary cells in the reference model is the same used for the unitary cells in the coil submodel and in the single tape submodel of the multi-scale models. The shape of the unitary cell in figure 2 applies to systems wound with (RE)BCO coated conductors, and the dimensions could depend on the wounding process. For different systems with different conductors, like the MgB$_2$ coil in [19], other unitary cells with different shapes are required.

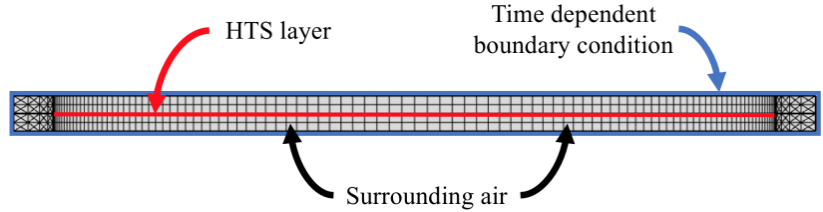

**Figure 2.** Unitary cell mesh. The red line represents the HTS layer, whose thickness has been increased in the drawing. The rest of the grey region represents the surrounding materials.

The system is simulated for one period of a sinusoidal transport current. Integral constraints are used to enforce the transport current in each tape [28]. Once the components of the magnetic field strength are computed, the average hysteresis losses can be obtained as follows,

$$Q = \frac{2}{T}\int_{T/2}^{T}\int_{\Omega} 2\pi r E_\varphi J_\varphi d\Omega dt . \qquad (4)$$

The reference model was implemented in COMSOL Multiphysics. 5.3 PDE mode application [34].

**Multi-scale models**

Two different models of the case study are presented, the difference between these models is the number of analyzed tapes. Both interpolation methods, ICDF and linear, were applied to approximate the *J* distributions. As it is going to be shown in the next section, the results obtained with the reference model in the previous section show that the most significant amount of hysteresis losses is located in the upper pancakes. The distribution of the losses is not uniform across the pancakes. At the inner and outer radius of the pancakes, the losses show a caving shape. To be able to reproduce this caving, the multi-scale models require an increased density of analyzed tapes in the innermost and in the outermost turns of the coil. It should be noted that the inner pancakes have a larger $J_c$ because the magnetic field is mostly parallel to the tape wide surface. Therefore, it produces a large non-penetrated region and lower losses. At the same time, the external pancakes exhibit a larger penetrated region mostly due to the larger amount of the magnetization currents. These last observations are consistent with the results reported in [16, 17].

The number and position of the analyzed tapes in each model are chosen to estimate the losses and current densities with the least impact on accuracy compared to the reference model for a reasonable computational speed and load. The best accuracy was then achieved by grading the distribution of the analyzed tapes so that more analyzed tapes were located at the inner and outer radius of the top pancakes. Fewer tapes were chosen for the inner pancakes, which typically show negligible losses compared to the top pancakes. The first multi-scale model uses the set of analyzed tapes {1, 5, 40, 76, 80} for the pancakes 3, 4 and 5, while the pancakes 1 and 2 have just one analyzed tape. The total number of analyzed tapes amounts to 17. When there is just one analyzed tape inside the pancake the *J* distribution and the losses in the analyzed tape are considered to be the *J* and the losses occurring in all the tapes in the corresponding pancakes. The second multi-scale model has 40 analyzed tapes, the set of analyzed tapes in every pancake is {1, 2, 5, 26, 55, 76, 79, 80}, this set once again respects the directive of an increased density of analyzed tapes in the innermost and in the outermost turns. For these models the convergence criterion ($\varepsilon$) for *Er* is 0.03.

The coil submodel as well as the single tape submodel were implemented in COMSOL Multiphysics. 5.3. Both interpolation methods were implemented in MATLAB®. The full method including the steps to import and export the data was implemented in COMSOL's LiveLink™ for MATLAB®.

## 4. Results

**Reference model results**

First, simulations are performed with the reference model to get what is considered for all the subsequent iterative models as the best estimation on the magnetic field, current density distributions and losses. The reference model and the rest of the models were simulated for two sinusoidal transport currents with different amplitudes, $I_m$=25 A and $I_m$=50 A, respectively, and 50 Hz frequency. For the first transport current ($I_m$=25 A), the normalized current density ($J_n=J/J_c$), the magnetic flux density magnitude (/$B$/) and the average losses are shown in the first row of figure 3. The plots for $J_n$ and |$B$| present the values at peak current ($t$=15 ms). The x-axis in the plot of the average hysteresis losses presents the tapes inside every pancake, numbered from inside the coil outwards. The losses plots in figure 3 show five lines, each one representing the losses in a different pancake. The pancake 1 is the pancake located at the center of the coil, while pancake 5 is located at the top. The losses in pancake 5 are approximately two orders of magnitude larger than the losses in pancake 1, this sensible difference is due to the higher current penetration of the tapes due to the higher magnetic field developed in the pancake 5, and also it is due to the difference in the field direction. Although there are slight increments in the losses at the extremities of a given pancake, these remain within the same order of magnitude for the whole pancake.

The results for the second transport current ($I_m$=50 A) are shown in the first row of figure 4. Once again, the losses in pancake 5 are approximately two orders of magnitude larger than those in pancake 1. This case was chosen to show the saturation of pancake 5. Under these conditions, the slight losses increment in the extremities of the pancakes became slight decrements for the two upper pancakes. Besides, as it can be seen in the text boxes in figures 4 and 5, the total losses for the transport current $I_m$=50 A, are more than 6 times larger than the losses when the transport current is $I_m$=25 A. The computation times required for the reference model are reported in table 3.

**Multi-scale models results and comparisons**

The results both interpolation methods are presented for the simulations with a transport current of amplitude $I_m$ =25 A. For the 50 A transport current, the coefficients of determination between the $J$ distribution of the reference model and the $J$ distribution of the multi-scale models with ICDF interpolation are equal in the first three significant figures to the coefficients of determination when linear interpolation is applied. Thus, it is redundant to present the results for both interpolation methods at $I_m$ =50 A, and just the results with ICDF interpolation are presented.

The results for the case of a transport current with an amplitude $I_m$ =25 A are presented in figure 3. Plots for $J_n$, |B| and the average hysteresis losses, are shown in the second and third row. The results of the reference model (black dashed lines) are also shown. It is important to clarify that the plots for $J_n$ and |B| show only the results from the ICDF interpolation of $J$. Indeed, any differences with the linear interpolation method cannot be appreciated at the given scale of the figures. The average losses when the ICDF interpolation is applied are plotted in red. The circles denote the results in the analyzed tapes, while the solid lines show the interpolated losses over all the tapes. The blue crosses correspond to the results in the analyzed tapes, when the linear interpolation is applied.

Figure 5 shows the results of the multi-scale model with 40 analyzed tapes when the uniform $J$ distribution is firstly applied and subsequently at iterations 1 and 3. No losses plot is presented for the initial uniform $J_0$ distribution since, at this point of the process, the single tape submodel has not yet been simulated. It is noteworthy that the iterative multi-scale method allows the $J$ distribution to quickly evolve from a uniform distribution to a more accurate $J$ distribution that takes into account shielding effects. Consequently, the average losses are converging to the losses obtained with the reference model. The central tapes and the tapes of the pancake 5 experience a faster convergence, this is due to the fact that the background field reproduced in the early iterations is sufficient for the single tape submodel to reproduce the losses. In the other hand, the external tapes of the pancake 1 require more iterations to accurately reproduce the losses. The evolution of the $J$ distribution is clearly exemplified in figure 6, where the $J$ distributions in the tape 80 of the pancake 5 for different iterations are shown together with the distribution of the reference model.

Figure 7 presents the relative errors in the total average losses and the coefficient of determination [35] in $J$ distributions calculated comparing the results at each iteration with the results of the reference model. The relative error in the total average losses is defined by

$$Eq_k = \left|\frac{Q_r - Q_k}{Q_r}\right|, \qquad (7)$$

where $Q_r$ is the total average losses of the reference model, while $Q_k$ is the total average losses of the multi-scale models at iteration $k$. The coefficient of determination is denoted by $R_k^2$, and is defined by

$$R_k^2 = 1 - \frac{\sum_{i=1}^{m}(J_i^r - J_i^{full\_k})^2}{\sum_{i=1}^{m}(J_i^r - \bar{J}^r)^2}, \qquad (8)$$

where $J_i^r$ is a vector containing the evenly sampled $J$ distribution of all the tapes, at all time steps, concatenated one behind the other, computed with the reference model. While $J^{full\_k}$ is a vector containing the interpolated $J$ distribution in all the tapes at iteration $k$. Both vectors contain $m$ samples. Both quantities, $Eq_k$ and $R_k^2$ are dimensionless quantities. The results when the linear interpolation is applied are presented in figure 7 a), while the results when the ICDF interpolation is used are shown in figure 7 b). For the linear interpolation, the convergence criterion ($\varepsilon=0.03$) is reached at iteration 10, and it is reached at iteration 11 for the ICDF interpolation.

The results for the transport current with an amplitude $I_m = 50$ A are presented in figure 4. The plots for $J_n$, $|B|$ and the average hysteresis losses are shown in their respective columns. As in figure 3, the results of the reference model are shown in the first row, while the results of the multi-scale models are shown in the second and third row. The evolution of the results from one iteration to other is shown in figure 8. For this transport current the convergence criterion is reached at iteration 8 for the model with 17 analyzed tapes, while it is reached at iteration 9 for the model with 40 analyzed tapes.

The data flow from the coil submodel to the single tape submodels is implemented through the field at the boundary of the analyzed tapes. This boundary field accounts for the effect of the rest of the tapes over the analyzed tape. At a given point, the local effects in the field produced by the $J$ distribution in one specific tape vanish when the distance between the given point and the specific tape is increased. Therefore, it is possible to achieve a satisfactory estimation of the total losses, and even of the local losses, with $J$ distributions with $R^2$ around 0.95. This last fact can be observed in the figures 7 and 8.

For the cases with $I_m = 25$ A, the relative losses error at the first iteration is around 0.55, and drastically decreases to less than 0.12 at the second iteration. The relative error of 0.55 give a clear idea of the error that can be incurred when the uniform $J_0$ distribution is considered. The next point of interest is that, for the case with $I_m = 50$ A, the estimation of the losses and the $J$ distribution are more accurate at the first iteration than for the case with $I_m = 25$ A. The most important observation is that, at iteration 5, the losses errors in all the cases are about 0.01 or less.

The two plots in figure 7 are qualitatively similar. Nevertheless, the accuracy regarding the $R^2$ in $J$ is slightly better when the ICDF interpolation is applied, the difference is found in the third significant figure, thus it is not visible at the given scale of the plot. This observation is consistent with the ICDF interpolation exhibiting a better behavior in the zones covering the moving fronts of current densities.

These zones of these drastic changes in the current density represent just a small part of the overall $J$ distribution in the tapes. This suggests that the election of the current density interpolation method does not have a significant impact in the global results.

For the tested conditions, the local losses for all the tapes of a given pancake are similar, thus having a good approximation of the losses in the central tape of the pancakes 1 and 2 provides satisfactory approximations of the losses in those pancakes. Of course, it is not possible to reproduce the variations in the extremes parts of pancakes 1 and 2, but since the losses in the pancakes 4 and 5 are one order of magnitude larger, the incurred errors are negligible. The simulations are conducted in order to validate the proposed models, even when the simulations consider zero external field conditions, the scope of the models is not limited to these conditions, and if it is required the external field can be considered.

**Computation time**

The computation times required to run the iterative multi-scale models, as well as the reference model are reported in table 3. The converge criterion is reached after a different number of iterations for each case. The number of iterations needed to reach the convergence criterion are also reported in the "total iterations" column in table 3. The computer used to perform the simulations is a MacBook (3 GHz Intel Core i7-4578U, 4 cores, 16 GB of RAM). The linear and ICDF interpolation processes need 35 s and 60 s, respectively. Even when the ICDF interpolation requires around twice the time of the linear interpolation, this difference is not significant when it is compared with the computation time required by a complete iteration. The average computation time required to run the single tape submodel is 65 s for t $I_m$=25 A, and 180 s for $I_m$=50 A. The computation time required to run the coil submodel is 750 s, these times include the time required to import the current density distribution and to export the magnetic field. It is important to point out that the times required by the coil submodels remain the same regardless of the transport current amplitude. Additionally, it is necessary to consider 200 s, which is the computation time required to run the coil submodel with uniform current density before the first iteration.

The time required by the multi-scale models to complete one iteration depends on the number of analyzed tapes. The times reported in table 3 are the times required to run the multi-scale models in series, all these times are less than the time required for the reference model. For the models with 40 analyzed tapes the computation time is around the 50 % of the time required for the reference model, while for the models with 17 analyzed tapes this quantity is around 27 %. It is possible to reduce the computation times if the convergence criterion is relaxed. For example, if $\varepsilon=0.2$, then the convergence is reached at iteration 4, while at the same time the relative losses error is around 0.01, as it can be seen in figures 7 and 8. The computation time required to run the multi-scale models up to the 4ª iteration are also reported in the las columns of table 3.

**Table 3.** Computation time.

| Model | $I_m$ [A] | An. tapes | Interpolation | Total Iter. | Comp. time [h] | % of Ref. model time | Comp. time at 4º iter. [h] | % of Ref. at 4º iter. |
|---|---|---|---|---|---|---|---|---|
| Multi-scale | 25 | 17 | ICDF | 11 | 5.7 | 30.5 | 2.2 | 11.7 |
| | 25 | 40 | ICDF | 11 | 10.3 | 55.1 | 3.8 | 20.3 |
| | 25 | 17 | Linear | 10 | 5.1 | 27.3 | 2.1 | 11.2 |
| | 25 | 40 | Linear | 10 | 9.3 | 49.7 | 3.7 | 19.8 |
| | 50 | 17 | ICDF | 8 | 8.5 | 23.7 | 4.3 | 12.0 |
| | 50 | 40 | ICDF | 9 | 19.9 | 55.4 | 8.9 | 24.8 |
| Reference | 25 | ---- | ---- | ---- | 18.7 | 100.0 | ---- | ---- |
| | 50 | ---- | ---- | ---- | 35.9 | 100.0 | ---- | ---- |

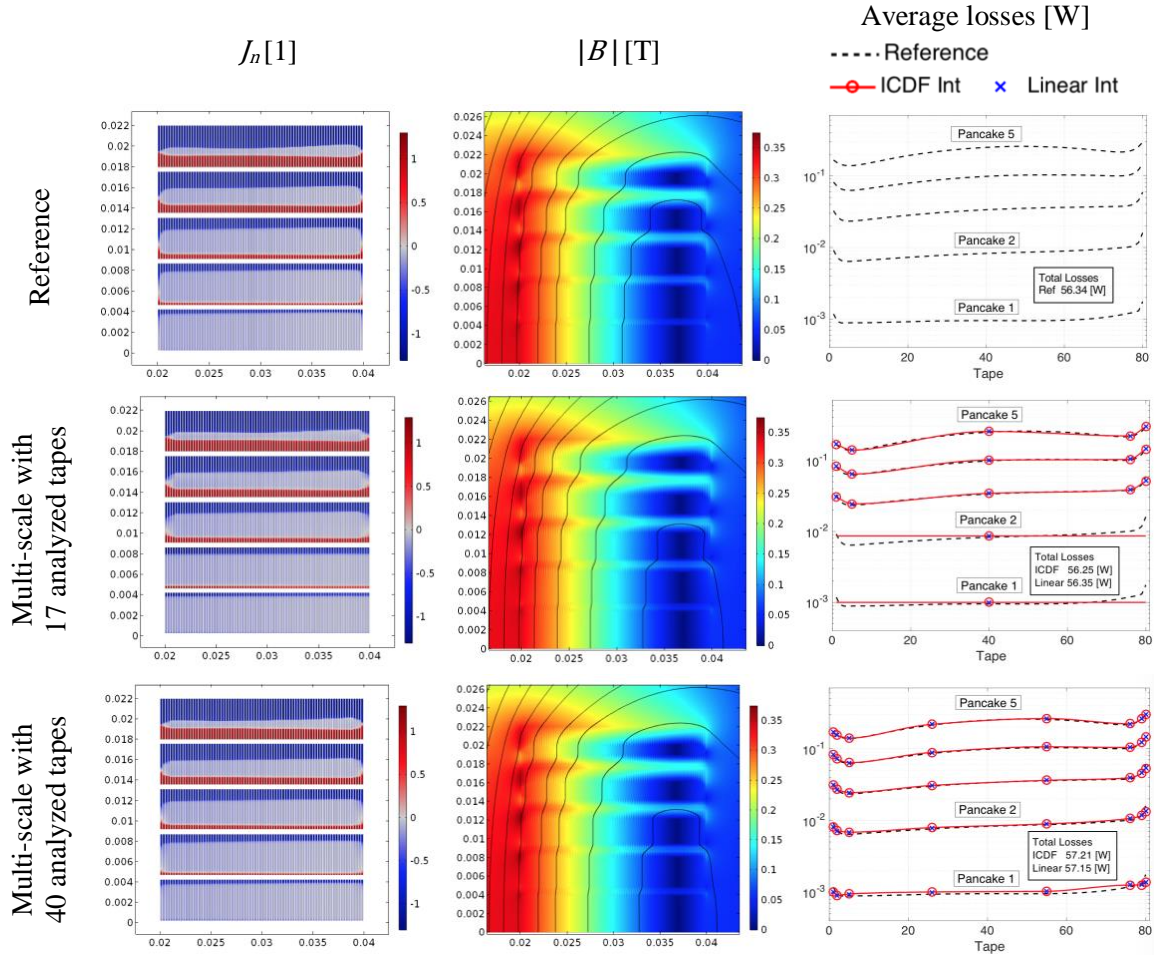

**Figure 3.** Reference and multi-scale models results with $I_m$=25 A. The first row shows the results of the reference model. The second and third row shows the results of the 17 and 40 analyzed tapes multi-scale models, respectively. The plots for $J_n$ and $|B|$ show the results at peak transport current, $t$ =15 ms.

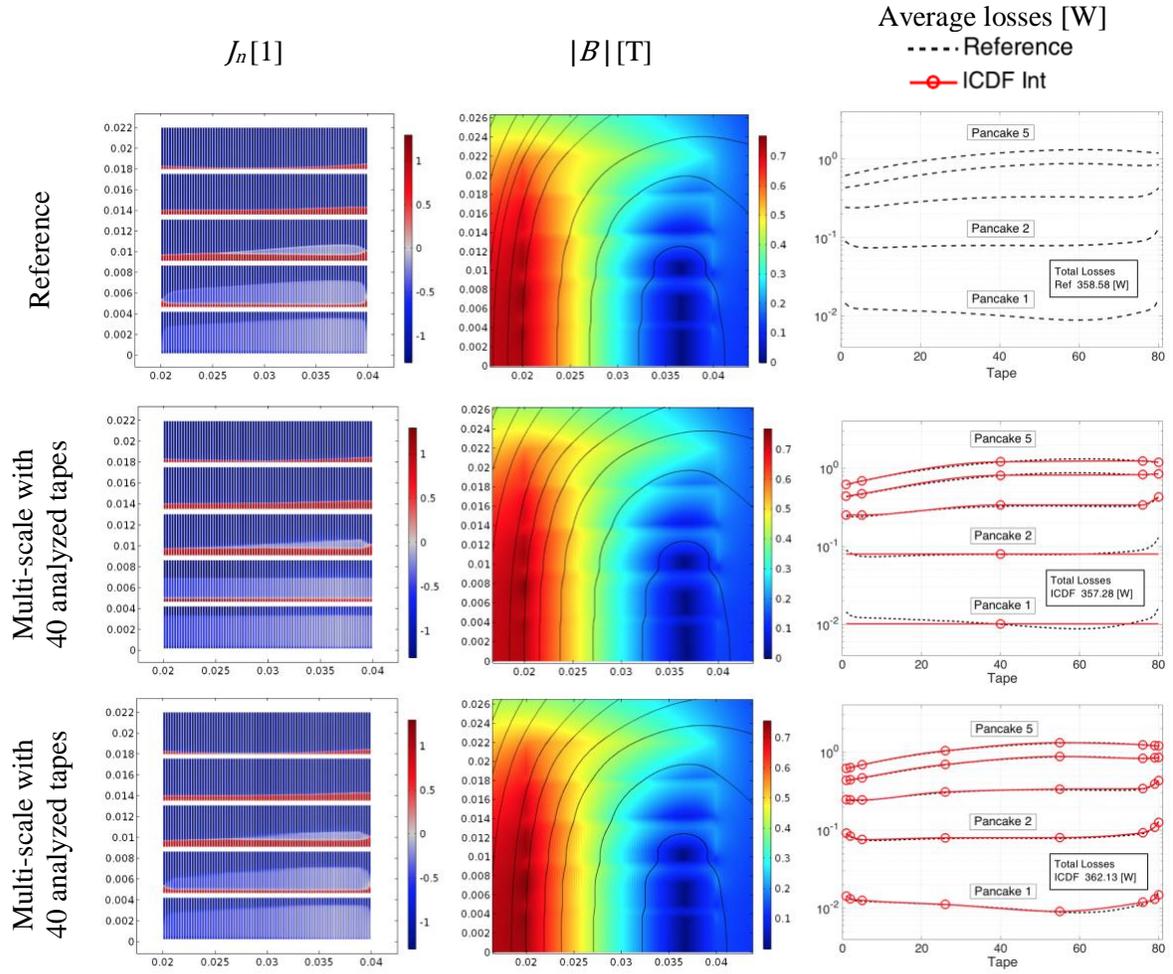

**Figure 4.** Reference and multi-scale models results with $I_m$=50 A. The first row shows the results of the reference model. The second and third row shows the results of the 17 and 40 analyzed tapes multi-scale models, respectively. The plots for $J_n$ and $|B|$ show the results at peak transport current, $t$ =15 ms.

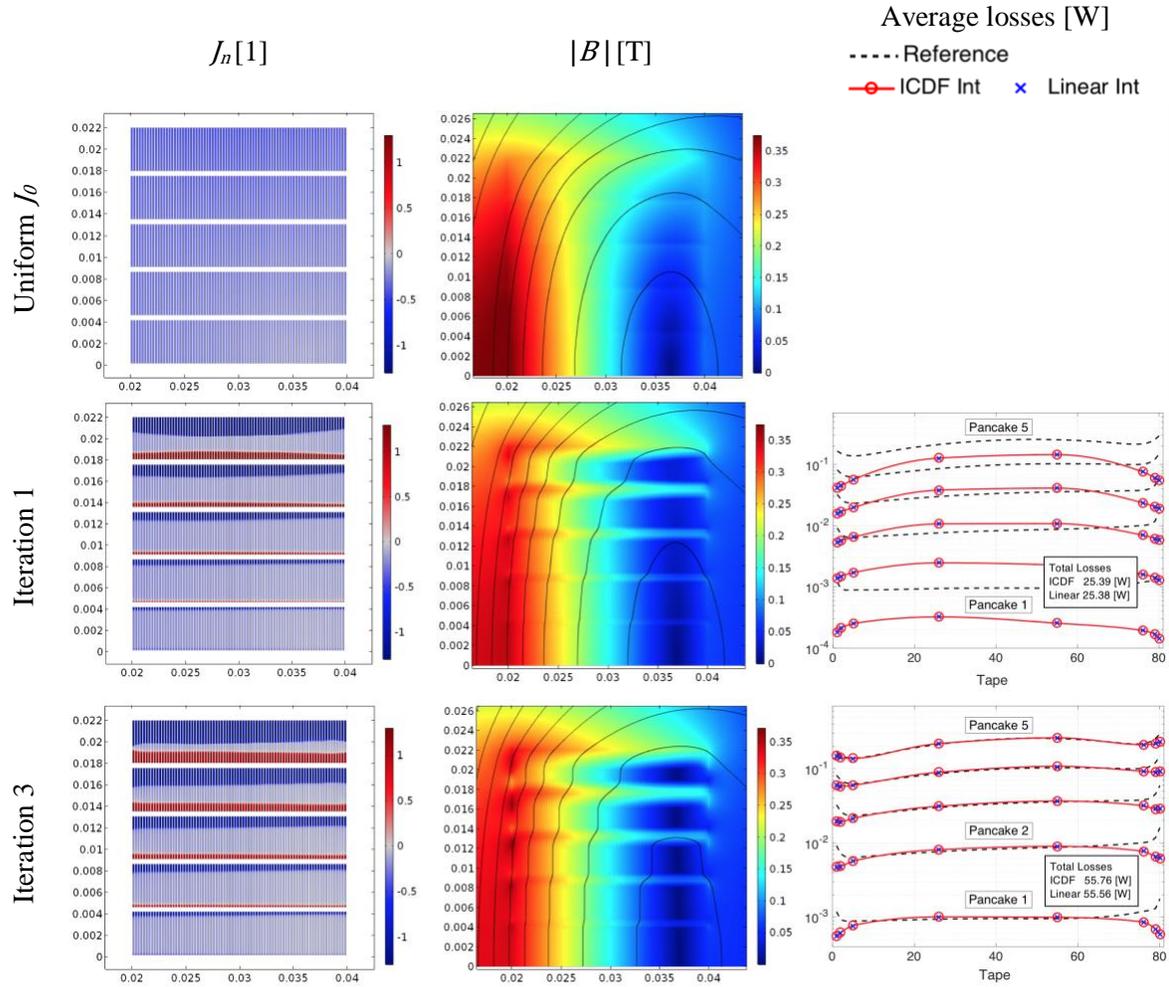

**Figure 5.** 40 analyzed tapes multi-scale model evolution with $I_m=25$ A. The first row shows the results when a uniform $J$ distribution is applied. The second and third row shows the results at iterations 1 and 3, respectively. The plots for $J_n$ and $|B|$ show the results at peak transport current, $t=15$ ms.

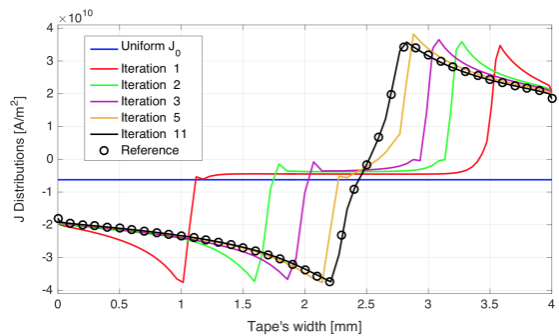

**Figure 6.** $J$ distributions in the tape 80 of the pancake 5, with $I_m=25$ A, at the peak of the transport current, $t=15$ ms. The distributions evolve from a uniform distribution to a distribution that is almost equal to the $J$ distribution of the reference model.

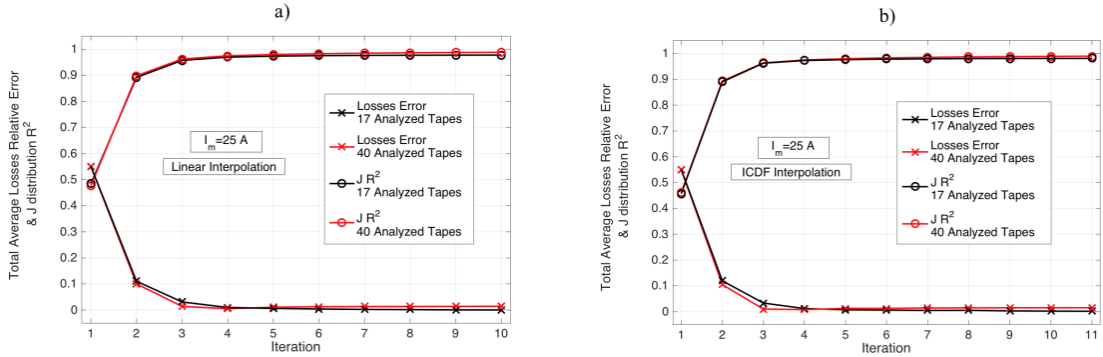

**Figure 7.** Total average losses relative errors and $J$ distribution $R^2$. Iterative multi-scale models with $I_m$=25 A.

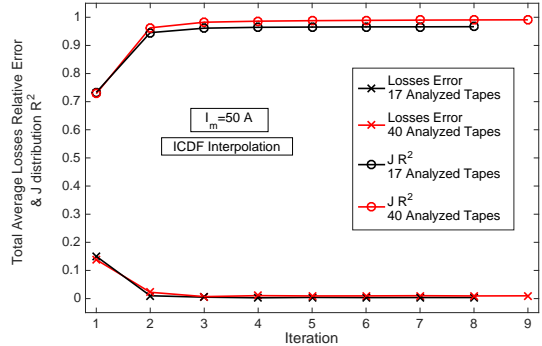

**Figure 8.** Total average losses relative errors and $J$ distribution $R^2$. Iterative multi-scale models with $I_m$=50 A.

## 5. Conclusion

In this manuscript, a new multi-scale method has been presented. This new approach relies on an iterative scheme to obtain a better $J$ distribution in the superconducting tapes of the system at every time step. Thus, a more accurate background field is computed that leads to a better estimation of the total losses of the system. This is a clear advantage over the classic multi-scale method since it provides an accurate $J$ and $B$ distribution over the entire system. There are important estimations when the quality of the field is also a matter of interest.

Since only a few tapes are analyzed, the information obtained in the analyzed tapes must be propagated to the remaining non-analyzed tapes. The linear interpolation of the average losses is a fairly straightforward process to get the losses in every tape. However, this method is unable to lead to the actually expected current densities in the non-analyzed tapes as explained in the appendix A. Therefore, a new interpolation method, referred to as ICDF, has been proposed here. However, it is useful mainly to get a more accurate current density and local magnetic field distribution, but it does not offer much improvement on the estimation of the losses.

The iterative multi-scale models are a useful tool for addressing the analyses of large-scale systems, for which it is not possible to deal with a *H*-formulation reference model. There is not a method to determine which is the best set of analyzed tapes. But, the unevenly distribution used here allows an accurate reproduction of the shielding currents, and consequently of the background field and the losses. The multi-scale models have the additional advantage that can be constructed with the objective of achieve an almost arbitrary level of local accuracy. Towards this end, it is necessary to increment the number of analyzed tapes in the specific region where the accuracy is required. In the cases where the size of the system makes it possible to deal with the *H*-formulation reference model, this reference model could be just necessary when local high accuracy results are required over the full system.

## Appendix A

The new interpolation technique is based on the inverse cumulative density function (ICDF) interpolation. This technique was initially developed to interpolate probability density functions as described in [24]. The ICDF interpolated distribution is defined by

$$\widehat{f_{2\_ICDF}}^{-1}(y) = (\alpha)\widehat{f_1}^{-1}(y) + (1-\alpha)\widehat{f_3}^{-1}(y),  \tag{A.1}$$

where the $\widehat{f_1}$ is the cumulative distribution function of $f_1$, given by

$$\widehat{f_1}(x) = \int_{-\infty}^{x} f_1(x')dx',  \tag{A.2}$$

the rest of the cumulative functions are defined in a similar way, and $\alpha \in [0,1]$.

The application of equation (A.1) is only possible when the cumulative functions are monotonically increasing functions, which in turns requires the original functions to be positive. One strategy proposed in [28], is to separate the functions into their positive, negative and average or constant components, and interpolate each individual component separately. It is important to note that for the negative components, the interpolation process is applied to the absolute value of these components. The constant components account for the transport current in the tapes. In this case the interpolation of the constant component is not necessary because the transport current is the same for all the turns connected in series in the coil.

The next step is to normalize the positive and negative components. The normalizing factor is the defined integral of the component, thus the normalized positive component is defined as

$$\overline{J_{1\_pos}} = \frac{J_{1\_pos}}{\langle J_{1\_pos} \rangle},  \tag{A.3}$$

where $J_{1\_pos}$ is the positive component of the distribution $J_1$ and the normalizing factor is defined as

$$\langle J_{1\_pos} \rangle = \int_q J_{1\_pos}(x) dx \tag{A.4}$$

and $q$ is the tape's width. Thus, the normalized components have a defined integral equal to 1.

The direct application of equation (A.1) to interpolate the normalized current density components produces the expected results when the positive and negative components have just one bump. For some functions, like functions with two or more bumps, the direct application of equation (A.1) may produce spurious bumps in between the bumps of the original functions. A similar problem with these spurious bumps, so-called "translating bumps", was addressed in [28]. The origin of these "translating bumps" is a computational issue, they do not represent any physical phenomenon, thus its presence should be avoided. The proposed solution in [36] is to use a multi-resolution scheme, which interpolates different band passed components of the original functions separately. The solution proposed in this manuscript is to add an offset $\delta$ before the application of equation (A.1). The defined integral of the normalized components plus the offset is equal to a constant $\beta > 1$, as follows:

$$\beta = \int_q \left( \widetilde{J_{1\_pos}}(x) + \delta \right) dx . \tag{A.5}$$

The offset $\delta$ was chosen, so that the value of $\beta$ is equal to 1.5. Equation (A.5) is also valid for the other normalized positive components or normalized absolute value of the negative components. Smaller values for the offset do not eliminate the translating bumps, while larger ones mask the original shape of the functions. The offset causes the cumulative function to be strictly increasing, avoiding step changes in the inverse cumulative function, which in turns avoids the translating bumps.

Now, the interpolated normalized components are obtained similarly to (A.1),

$$\widetilde{J_{2\_pos}}^{-1}(y) = (\alpha)\widetilde{J_{1\_pos}}^{-1}(y) + (1-\alpha)\widetilde{J_{3\_pos}}^{-1}(y) , \tag{A.6}$$

where

$$\widetilde{J_{1\_pos}}(x) = \int_{-\infty}^{x} \left( \widetilde{J_{1\_pos}}(x') + \delta \right) dx' . \tag{A.7}$$

The rest of the cumulative functions are defined in a similar way. In order to retrieve the interpolated component, the offset needs to be subtracted, and the interpolated normalized component must be denormalized. Thus, the interpolated positive component is defined as

$$J_{2\_pos} = \{(\alpha)\langle J_{1\_pos} \rangle + (1-\alpha)\langle J_{3\_pos} \rangle\} \overline{J_{2\_pos}} . \tag{A.8}$$

Finally, the interpolated *J* distribution is found by combining the average, positive, and negative interpolated components.

Figure A1 shows an example of the ICDF interpolation, as described here. The distributions $J_1$ and $J_3$ are used to approximate the distribution $J_2$. The distributions $J_1$ and $J_3$ describe the current density as a function of the tape's with, at a given time step. Each distribution represents a different tape inside the case study and were obtained from the reference model.

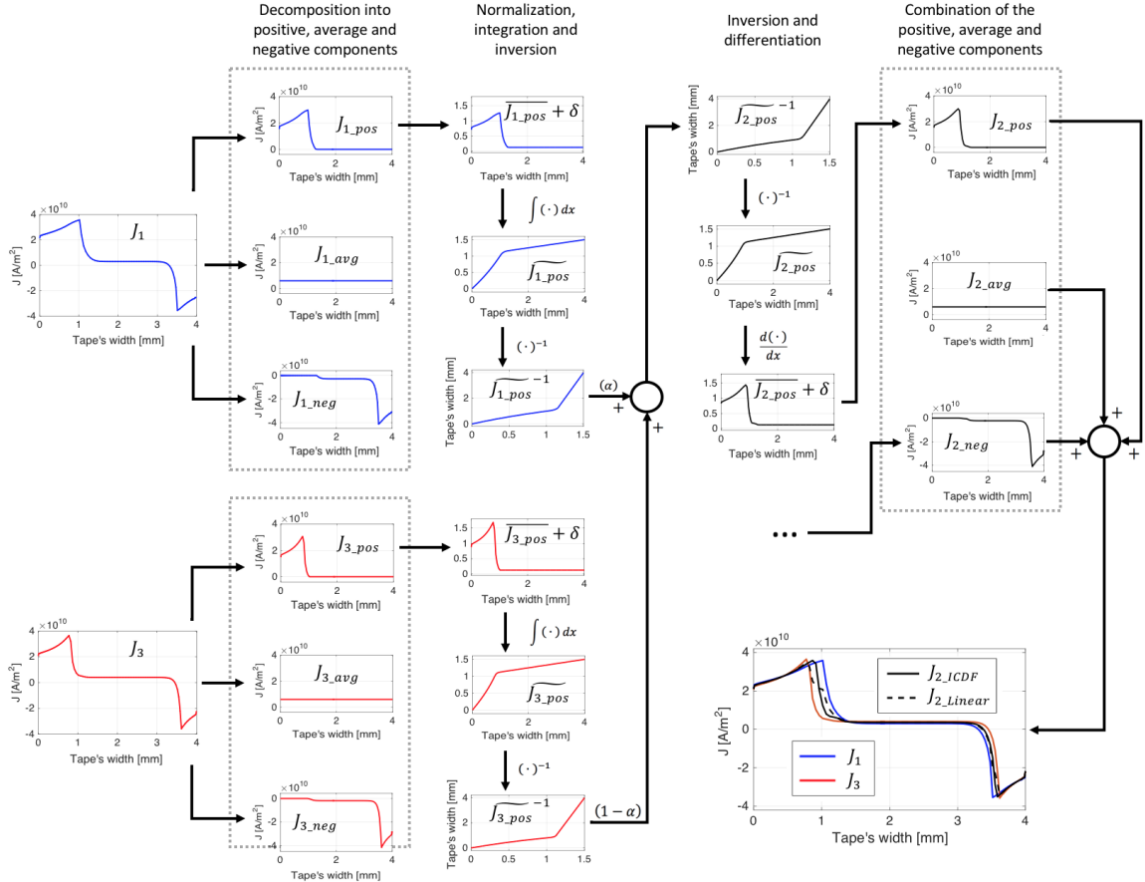

**Figure A1.** ICDF interpolation of two *J* distributions, $J_1$ and $J_3$ are used to approximate $J_2$. The distributions $J_1$ and $J_3$ were obtained from the reference model, they are shown in the first column. The second column shows the decomposition of the distributions into its positive, average and negative components. The third column shows the normalization, integration and inversion of the respective positive components. Between column 3 and 4 the $\widetilde{J_{1\_pos}}^{-1}$ and $\widetilde{J_{3\_pos}}^{-1}$ are interpolated to find $\widetilde{J_{2\_pos}}^{-1}$. The fourth column shows the reverse process of the third column, this is the inversion and derivation of the interpolated component. The fifth column shows the positive, average and negative interpolated components, the process to interpolate the negative component is similar to that of the positive component. The linear interpolation produces the averaging of the original *J* distributions. While the ICDF interpolation produces the displacement of the current density fronts.